May 7, 2014

# Milne Quantum-Universe Redshift-Luminosity Correlation

Geoffrey F. Chew

*Theoretical Physics Group*
*Physics Division*
*Lawrence Berkeley National Laboratory*
*Berkeley, California 94720, U.S.A.*

Abstract

Milne's classical homogeneous-universe cosmology predicts a product of Hubble constant with luminosity distance that equals $z + z^2/2$, where $z$ is redshift. Supernova-data are consistent with this relation, supporting *quantum*-theoretic considerations that reveal Milne's universe as 'non-empty'.

**Introduction**

Milne's classical cosmology, [1] interpreted by general relativists as corresponding to zero universe energy, was dismissed during the last century by all save its inventor. [2] But two recent developments, one experimental and one theoretical, are resurrecting Milne's *alternative* to general relativity (GR). (1) Data from supernovae at redshifts approaching $z = 1$ disagree *qualitatively* with GR expectations. [3] (2) A heretofore unutilized unitary Hilbert-space Lorentz-group representation has been found to define a positive-energy 'Milne quantum universe' (MQU). [4] We here derive, in a homogeneous approximation to MQU, a simple relation between redshift and luminosity distance that is consistent with current data.

Although his cosmology has been regarded a special case of the Friedman-Robertson-Walker metric, [5] Milne did not follow that route; neither did the present author. For different reasons we have *both* employed the Lorentz group as *foundation* for 'reality within spacetime'. Milne's cosmology--pure 'classical kinematics'--was Lorentz-group and spacetime motivated--with reference neither to quantum theory nor to gravity and energy. The present author has invoked a 'bundle' of 3-dimensional Milne negatively-curved metricized base space, a 3-dimensional unmetricized fiber space *and* a Hilbert space. This bundle *cosmologically* satisfies Dirac's (quantum) principles through previously-unutilized *unitary* Hilbert-space representations of the Lorentz group. Gravity, with energy as source, is represented together with electromagnetism.

After achievement of definition for 'classical reality' in a gravity-encompassing single-universe quantum-cosmological theory, [4] we have begun to contemplate experimental tests. Because the spacetime meaning to which quantum theory has led us is that of Milne, we are exploring the natural possibility that Milne's cosmology is a Hubble-scale classical homogeneous approximation to *our* quantum theory of the universe—a theory, more general than his, which deals with *all* scales between Planck's and Hubble's.

*Beyond* assignment of foundational status to the Lorentz group, our theory of spacetime reality recognizes *two* foundational *integers*, one 'large' and one 'huge', that allow *different approximations* to be 'physically viable', *separately*, for *different* limited-scale ranges. Each such approximation within its scale range is, for all practical physics purposes (FAPPP), 'reliable'. (Our large integer associates with the reciprocal of a 'GUT-scale fine structure constant' and our huge integer with the universe's total energy. [4]) Although reliability of Milne (classical) Hubble-scale cosmology remains to be deduced from our quantum cosmology, the supernovae data has provoked the present paper.

**Milne Spacetime**

Milne spacetime occupies the interior of a forward lightcone, with the 'age' of any spacetime point equal to its Minkowski distance from the lightcone vertex. Spacetime was seen by Milne as 'filled' with metricized 3-dimensional (noncompact) hyperbolic (curved 3-space) manifolds, each belonging to a *single* (positive) age. Redshift immediately follows, with Hubble's 'constant' (*H*) *equal* to the reciprocal of universe age ($c = 1$).

The *curvature* of a Milne 3-manifold equals *H*--independently of location within the manifold. Euclidean ('flat') geometry is asymptotically approached in the limit as age approaches infinity. 'Milne relativity', distinct from either special or general relativity (and not only more general than his Hubble-scale classical cosmology but extendable to quantum cosmology), implies that any two universes related by a *global* Poincaré transformation are the *same* universe (*not* 'alternative' universes). *MQU* enjoys 'fixed and settled reality'.

A Milne-Lorentz *boost* shifts spatial locations at a fixed age. To any spacetime location there associates a continuous (labeled by 3 Euler angles) set of rotationally-related 'local frames'. In any local frame the positive-timelike 4-vector displacement from the lightcone vertex of the location in question has components ($\tau$, 0, 0, 0), where $\tau$ is the location's age.

Let the symbol $\beta$ denote the dimensionless positive 'boost distance', along a hyperboloid geodesic, between two spacetime points of the same age. If $c = 1$ the 4-vector spacetime location of one of these points, in any of the (rotationally-related) local frames belonging to the *other* point, is $\tau \times (cosh\,\beta, \boldsymbol{n}\,sinh\,\beta)$, where $\boldsymbol{n}$ is a unit 3-vector whose pair of direction ('polar') coordinates refer (for 'origin') to the orientation of the *other* point's local frame. Spacetime points of *different* age but *parallel* location 4-vectors share the *same* 3-vector $\boldsymbol{\beta} \equiv \beta\boldsymbol{n}$. (They occupy the *same* location in 'boost space'.) Thus Milne spacetime is coordinated by $\tau$, $\boldsymbol{\beta}$ once some 'origin' within a 6-dimensional manifold—the product of a (compact) 3-sphere with a (noncompact) 3-hyperboloid--has been designated. [4]

The 3-hyperboloid Lorentz-invariant (*dimensionful*) metric is

$$(ds)^2 = \tau^2 \{ (d\beta)^2 + sinh^2\beta\, [(d\theta)^2 + sin^2\theta\, (d\varphi)^2] \}, \qquad (1)$$

where $\theta$ and $\varphi$ are polar coordinates specifying the direction $\boldsymbol{n}$. The (4-spacetime) Minkowski metric is the sum of two *separately*-invariant terms: $(d\tau)^2 - (ds)^2$. Along any temporally-forward *lightlike* trajectory, $ds = d\tau$. Our theory supposes *Hubble-scale* light propagation to follow approximately such a 'Milne trajectory' (which ignores sub-Hubble-scale matter clumping—regarding matter as uniformly distributed).

For supernovae with redshifts of order 1 that share a common ('standard') energy release, the supposition that both supernova 'sources' and telescope 'sinks' are 'almost at rest' ($v/c \sim 10^{-3}$) in their respective local frames allows the straightforward computation in the following section of an unambiguous relation (no arbitrary parameter) between redshift and 'luminosity distance'.

**Hubble-Scale Milne Relation between Redshift and Luminosity Distance**

For boost-distance $\beta$ between light source at age $\tau_{source}$ and light 'sink' at (later) age $\tau_{sink}$, an immediate consequence of light propagation according to $ds = d\tau$ is

$$\tau_{sink}/\tau_{source} = e^{\beta}. \qquad (2)$$

The ratio (2) *also* equals that between time *intervals* of energy emission and absorption in respective local frames.

The sink-source age ratio (2) further yields the ratio between emitted-photon (source frame) energy and absorbed-photon (sink frame) energy—i.e.,

$$e^{\beta} = 1 + z, \qquad (3)$$

where $z$ is the (standard) redshift parameter.

The definition of 'luminosity distance' [5] is

$$d_L \equiv (L/4\pi\ell)^{1/2}, \qquad (4)$$

$L$ denoting total energy emitted per unit time in source frame, while the symbol $\ell$ denotes a ratio

$$\ell \equiv P/A, \qquad (5)$$

the symbol $P$ representing power received (energy per unit sink time) by a mirror of area $A$ whose surface is perpendicular to light-propagation direction. We now show that, for light propagating along Milne geodesics,

$$d_L = \tau_{sink}\, e^{\beta} \sinh \beta, \qquad (6)$$

or, equivalently, $Hd_L = z + z^2/2$, once $\tau_{sink}^{-1}$ is identified with Hubble's 'constant' $H$ and Formula (3) is employed to replace $\beta$ by $z$.

Suppose the mirror to be circular, with radius $b$. The metric (1) then, by Formula (7) below, relates $b$ to the tiny angle $\theta$ subtended in source frame by two geodesics that intersect at source, one geodesic passing through mirror center and the other contacting mirror perimeter:

$$b = \tau_{sink}\, \theta \sinh \beta. \qquad (7)$$

The mirror is reached by a tiny *fraction*, equal to $(\theta/2)^2$, of the total number of emitted photons. It follows from (2) and (3) that

$$(\theta/2)^2 L = e^{2\beta} P. \qquad (8)$$

Because the mirror (sink-frame) area $A$ is $\pi b^2$, $P = \ell \pi b^2$. The central result (6)—the motivation for this paper--follows from Formulas (4), (5), (7) and (8).

**Concluding Remarks**

Elsewhere-detailed electro-gravitational quantum dynamics proceeds through a Schrödinger equation whose Hamiltonian gravitational potential energy is proportional to the energies of 'MQU constituents'. [4] (The 'fixed and settled' energy-momentum current density is a classical second-rank symmetric-tensor gravitational potential's Dalembertian, divided by $G$.)

Central to Milne spacetime is an 'age arrow' that accompanies redshift. Milne's arrow of global time permits *temporally-stable* clumping of positive energy into galaxies. (Age arrow breaks Standard-Model CPT symmetry at galactic and Hubble scales--huge compared to those of particle physics.) Following Milne's thinking, we conjecture homogeneity of matter at *super*-galactic while *sub*-Hubble scales--smaller than that of the entire universe. In early-universe evolution, sub-Hubble-scale density inhomogeneities are presumed to have been electro-gravitationally generated.

In a private communication to the author, J. Finkelstein has pointed out that Milne's (Hubble-scale) cosmology is formally equivalent to an 'empty', $\Omega_\Lambda = \Omega_M = 0$, FRW universe—with zero cosmological constant proportional to $\Omega_\Lambda$ and zero matter density proportional to $\Omega_M$.

Efforts to base quantum cosmology on radiation-field Fock-space operators have led others to associate 'cosmological constant' with Fock-space vacuum energy. The author's quantum cosmology, although including electromagnetic and gravitational radiation within its fixed and settled classical reality, [4] has *no* radiation-field *operators* and its Fock space *lacks* a 'vacuum state'.

Acknowledgement

Lengthy discussion with Henry Stapp has been invaluable to this paper's preparation. Assistance has also been received from Don Lichtenberg, Saul Perlmutter and Ramamurti Shankar. Berkeley Chew has provided editorial advice.